\documentclass[%
 reprint,
 floatfix,
 superscriptaddress,
 amsmath,amssymb,
 prd,
 twocolumn
]{revtex4-2}

\usepackage{graphicx}
\usepackage{dcolumn}
\usepackage{bm}

\usepackage[normalem]{ulem}
\usepackage{hyperref}
\usepackage{mathrsfs}
\usepackage[caption=false]{subfig}
\usepackage[dvipsnames]{xcolor}
\definecolor{myred}{rgb}{0.89,0.259,0.204}

\usepackage{txfonts}
\newcommand{\comment}[1]{}

\begin{document}

\title{Relativistic scalar dark matter drag forces on a black hole binary}

\author{Shuo Xin}
\affiliation{SLAC National Accelerator Laboratory, Stanford University, Stanford, CA 94309, USA}
\affiliation{Physics Department, Stanford University, Stanford, CA 94305, USA}
\author{Elias R. Most}
\affiliation{TAPIR, Mailcode 350-17, California Institute of Technology, Pasadena, CA 91125, USA}
\affiliation{Walter Burke Institute for Theoretical Physics, California Institute of Technology, Pasadena, CA 91125, USA}

\begin{abstract}
    Dark matter around black holes can induce drag forces through dynamical friction and accretion, potentially affecting the orbital evolution and gravitational wave emission of binary systems. While dynamical friction from scalar field dark matter has been studied in the relativistic regime for single black holes, the case of a binary black hole (BBH) has remained unexplored. As a first step, we present a series of two-dimensional general-relativistic simulations of BBH in a wind tunnel for an asymptotically homogeneous scalar field background. We extract the drag forces, torque, mass and charge accretion acting on the binary, and analyze their dependence on the binary separation, velocity and the scalar field parameters. We find that the binary's drag is not a simple superposition of two isolated black holes; the presence of a companion modifies the gravitational wake and yields significant nonlinearities. This additional force and torque can (in principle) modify the inspiral and induce a dephasing of the gravitational wave signal. 
\end{abstract}
\maketitle

\section{Introduction}

Strong gravitational environments surrounding black holes (BHs) {have the potential} to offer profound insights into fundamental physics. The study of strong gravity has gained unprecedented momentum with recent breakthroughs: the Event Horizon Telescope's historic imaging of supermassive BH shadows\cite{eht}, the growing catalog of binary BH (BBH) merger events by LIGO/Virgo/KAGRA\cite{ligo, a_ligo, virgo} collaborations, and rapid technological advances toward next-generation facilities. These include ground-based interferometers like the Einstein Telescope and Cosmic Explorer\cite{Punturo:2010zz, Reitze:2019, Kalogera:2021bya}, alongside planned space missions LISA, TianQin, and Taiji\cite{lisa_org, eLISA:2013xep, Luo:2016, Luo:2020}, which will extend our observational reach across unprecedented mass ranges and distances.

{One important question is the impact of environmental effects on black hole binaries, which  using space-based gravitational wave detectors may be detectable for stellar mass BBHs in disk of active galactic nuclei, and (to a lesser extent) in environments of massive BBH \cite{chen2025muffled}.}
BHs moving through surrounding matter can experience forces analogous to the friction~\cite{Chandrasekhar1943} and Bondi~\cite{BondiHoyle1944,Bondi1952} accretion in fluid dynamics, arising purely from gravitational interactions. In particular, the BH also experiences dynamical friction, a drag force opposite to its direction of motion, due to gravitational interaction with surrounding matter~\cite{PhysRevD.104.103014,PhysRevD.108.L121502,clough2021continuity,Bamber:2022pbs}. Also, a spinning BH moving through a dark matter (DM) medium feels a lateral ``spin-curvature'' force perpendicular to its velocity and spin axes \cite{PhysRevD.98.024026,Wang:2024cej}. These environmental forces could alter the trajectories of inspiraling BHs and imprint observable dephasing in gravitational wave signals when the inspiral occurs in a dense matter environment \cite{PhysRevD.96.083017}. Although under typical galactic DM densities the effects are expected to be tiny \cite{PhysRevLett.93.061302}, various mechanisms could significantly enhance DM densities around BHs, including superradiant instabilities that create bosonic clouds bound to spinning BHs \cite{PhysRevD.85.044043, richard2015superradiance,
arvanitaki2015discovering,arvanitaki2017black,
frolov2018massive,dolan2018instability,siemonsen2020gravitational}, as well as accretion of DM into steep-density ``spikes'' around isolated BHs or BBHs \cite{Sadeghian:2013}, motivating a detailed study of their influence on BH dynamics.

Light bosonic scalar particles are a well-motivated extension of the Standard Model \cite{svrcek2006axions, arvanitaki2015discovering} (see \cite{Graham:2015, Marsh:2015xka} for reviews). They are a compelling candidate for DM \cite{Preskill:1983, Abbott:1983, Dine:1983, brito2017gravitational,Jungman:1996, Bertone:2005, Bergstrom:2000}. In such scenarios, their accretion onto BHs may result in the formation of distinctive DM overdensities (or ``clouds'') around the BHs \cite{Gondolo:1999, Sadeghian:2013, Eda:2013}. Furthermore, superradiant instabilities can give rise to gravitationally bound scalar clouds around spinning BHs by amplifying small field fluctuations \cite{Arvanitaki:2010, arvanitaki2017black, baryakhtar2017black, Rosa:2012, Pani:2012, Konoplya:2006}, even if the scalars constitute only a fraction of the DM \cite{richard2015superradiance}.

In recent years, there have been abundant analytic studies on dynamical friction in dark matter (DM) environments. This is motivated by scenarios in which compact objects (COs) like black holes or neutron stars encounter halos of DM, including both cold collisionless particles and wave-like bosonic fields. Analytical and semi-analytical studies have explored dynamical friction in ultralight (fuzzy) dark matter \cite{BarOr2019,Lancaster2020} and superfluid dark matter \cite{Berezhiani2019} models. 
Other works have examined dynamical friction in self-gravitating bosonic condensates and Bose-Einstein condensate dark matter, including the impact of self-interactions
\cite{Hartman2021}. The gravitational drag in the presence of bosonic matter clouds or boson stars has also been investigated \cite{Annulli2020a,Annulli2020b,Vicente2019}.

General-relativistic simulations can now shed new light on these scenarios. Traykova \emph{et al.} carried out simulations of a black hole moving through a cloud of scalar dark matter~\cite{Traykova:2021,Traykova:2023letter}. Their results demonstrated relativistic modifications to the drag force and highlighted the contribution of matter accretion (Bondi-type accretion) to the net force at high velocities. {For spinning black holes, additional forces akin to an aerodynamic Magnus force leading to a lateral deflection appear \cite{Dyson:2024qrq,Wang:2024}, although the sign may be different between scalar field and fluid matter \cite{Kim:2024zjb}.}

Motivated by these possibilities, in this work we examine the interaction between scalar dark matter and binary black hole systems through general relativistic simulations. In the case of active galactic nuclei disks, the influence on BBHs have recently extensively been studied in the hydrodynamic regime \cite{Li:2022eup,Li:2022pnc,Li:2023gyv,Dittmann:2023sha,Dittmann:2025aup}. 
We study the dynamical friction force and torque on BBHs immersed in a homogeneous scalar field propagating in a relativistic velocity ($v \sim 0.2c$-$0.7c$). Using numerical simulations, we extract the drag force and torque experienced by the BHs due to their interaction with the scalar field, as well as the momentum, angular momentum, mass and charge accretion onto the BHs. This extends previous studies of single BHs in scalar DM to the case of a binary system. The paper is organized as follows: in Sec.~\ref{sec:formulation} we outline our theoretical formulation and describe the numerical methods, in Sec.~\ref{sec:results} we present the results of our simulations, and in Sec.~\ref{sec:discussion} we discuss the implications and provide our conclusions. Throughout the paper, we use geometrized units where $c=G=1$.


\section{Methods}
\label{sec:formulation}

The spacetime metric is decomposed into spatial and temporal parts using a 3+1 decomposition \cite{gourgoulhon2007}
\begin{equation}
\begin{aligned}
    ds^2 = & - \alpha^2 dt^2 + \gamma_{ij} (dx^i + \beta^i dt) (dx^j + \beta^j dt)\\
    =& - \bar\alpha^2 d\bar t^2 + \bar\gamma_{ij} (d\bar x^i + \bar\beta^i d\bar t) (d\bar x^j + \bar\beta^i d\bar t),
\end{aligned}
\end{equation}
where the ($t,x,y,z$) is our simulation frame and ($\bar t,\bar x,\bar y,\bar z$) is the center-of-mass rest frame of BBH. 
The spacetime decomposition introduces a future-directed unit normal vector to each spatial slice, $n^\mu = (1/\alpha,-\beta^i/\alpha)$, where $\alpha$ are the lapse, and $\beta^i$ the shift, and $\gamma_{ij}$ is the induced spatial metric.
The simulation frame is obtained by first Lorentz boosting the rest frame and then a Galilean transformation $x\rightarrow x-vt$ so that the center-of-mass does not move in the simulation frame. The two frames are related by \cite{PhysRevD.104.103014,PhysRevD.108.L121502,Wang:2024cej},
\begin{equation}
    \bar t = t/\gamma - \gamma vx,\quad \bar x = \gamma x,
\end{equation}
leading to a transformation among the metric functions,
\begin{align}
\label{eq_metric_trans_op}
    \beta_x = &\,\, \bar \beta _{\bar x} + v (\bar \alpha^2 - \bar \beta^2  ), \\
    \beta_y = &\,\, \bar \beta_{\bar y} /\gamma,\quad \beta_z =  \bar \beta_{\bar z} /\gamma, \\
    \gamma_{xx} = &\,\, \gamma^2 \bar\gamma_{\bar x \bar x}-2\gamma^2 v \bar \beta_{\bar x} + \gamma^2 v^2 (-\bar \alpha^2 + \bar \beta^2 ),\\
    \gamma_{xy} = &\,\, \gamma \bar \gamma _{\bar x \bar y} - \gamma v \bar \beta _{\bar y}, \\
    \gamma_{yy} = &\,\, \bar \gamma_{\bar y \bar y}, \quad \gamma_{zz} =  \bar \gamma_{\bar z \bar z},
\label{eq_metric_trans_ed}
\end{align}
where $\bar \beta^2  = \bar \beta^{\bar i}\bar \beta^{\bar j}\bar \gamma_{\bar i \bar j}$. Explicit expressions for $\alpha$ and $\beta^i$ in terms of bar-ed quantities are provided in the {Supplemental Material}. In our numerical implementation, we obtain $\beta_i$ and $\gamma_{ij}$ using analytic expressions above, solve for $\beta^i$ from $\beta_j = \beta^i \gamma_{ij}$, and obtain $\alpha$ by

\begin{equation}
    \alpha^2 = \beta^i \beta_i - (-\bar \alpha^2 + \bar \beta^2)/\gamma^2.
\end{equation}

As an example, consider a single Schwarzschild BH in isotropic coordinates

\begin{equation}
    ds^2 = - \bar \alpha_{\rm iso}^2 d\bar t^2 + \bar \Psi_{\rm iso}^4 (d \bar x^2 + d \bar y^2 + d \bar z^2),
\end{equation}
where
\begin{equation}
    \bar \alpha_{\rm iso}^2 = \left( \frac{1-\frac{M}{2 \bar r}}{1+\frac{M}{2 \bar r}} \right)^2, \quad \bar \Psi_{\rm iso} = 1 + \frac{M}{2 \bar r}, \quad \bar r ^2 = \gamma^2 x^2 + y^2 + z^2.
\end{equation}
By Eq. \ref{eq_metric_trans_op}-\ref{eq_metric_trans_ed}, in the boosted frame we have $ \gamma_{xx} = \gamma^2 \bar\Psi^4_{\rm iso} - \gamma^2 v^2 \bar\alpha^2_{\rm iso} $, $\gamma_{yy} = \gamma_{zz} = \bar\Psi^4_{\rm iso}$, $\beta_x = v \bar \alpha^2_{\rm iso}$, $\beta^x = v\bar\alpha^2_{\rm iso} / (\gamma^2 \bar\Psi^4_{\rm iso} - \gamma^2 v^2 \bar\alpha^2_{\rm iso})$ and $\alpha^2 =\bar\Psi^4_{\rm iso} \bar\alpha^2_{\rm iso} / (\gamma^2 \bar\Psi^4_{\rm iso} - \gamma^2 v^2 \bar\alpha^2_{\rm iso})$, in agreement with Ref. \cite{Traykova:2021dua}.\\
    
\subsection{Binary black hole spacetime}
The BBH spacetime in the center-of-mass frame $(\bar t, \bar x, \bar y, \bar z)$ is obtained by the extended conformal thin sandwich (XTCS) method \cite{Papenfort:2021hod} which we summarize in this section.  

The metric in center of mass frame is assumed to be conformally flat
\begin{equation}
    \bar \gamma_{\bar i \bar j} = \bar \Psi^4 \delta_{\bar i \bar j},
\end{equation}

Let $\tilde D_{\bar i}$ denote covariant derivative with respect to conformally related metric (here simply the flat metric). The Hamiltonian and momentum constraints give rise to the following differential equations for $\bar \Psi, \bar \alpha, \bar \beta^{\bar i}$:

\begin{align}
    \tilde D^2 \bar \Psi = & -\frac 18 \bar\Psi^{-7} \hat A_{\bar i \bar j}\hat A^{\bar i \bar j} - 2 \pi \bar \Psi^5 E, \\
    \tilde D^2 (\bar \alpha \bar \Psi ) = &\frac 78 \bar \alpha \bar \Psi^{-7} \hat A_{\bar i \bar j}\hat A^{\bar i \bar j} + 2\pi \bar \alpha \bar \Psi^5 (E+2S),\\
    \tilde D^2 \bar \beta^{\bar i} = & -\frac 13 \tilde D^{\bar i} \tilde D_{\bar j} \bar \beta ^{\bar j} + 2 \hat A^{\bar i \bar j} \tilde D_{\bar j} (\bar \alpha \bar \Psi^{-6}) + 16\pi \bar \alpha \bar \Psi^4 J^{\bar i}
\end{align}
where $\hat A ^{\bar i \bar j} = \frac{\bar \Psi^6}{2\bar \alpha }\left( \tilde D^{\bar i} \beta^{\bar j} + \tilde D^{\bar j} \beta^{\bar i} - \frac 23 \delta^{\bar i \bar j} \tilde D_{\bar k} \bar \beta^{\bar k} \right) $ is the traceless part of extrinsic curvature, and we adopt maximal slicing, setting the trace of the extrinsic curvature to zero.

The boundary conditions at infinity are
\begin{align}
    \lim_{\bar r \rightarrow \infty} \bar \alpha = \,\, 1,~
    \lim_{\bar r \rightarrow \infty} \bar \Psi = \,\, 1,~
    \lim_{\bar r \rightarrow \infty} \bar \beta^{\bar i}  = \,\,  \omega \hat \varphi^{\bar i},
\end{align}
so that the frame is corotating with the binary along the azimuthal $\hat \varphi$ direction with angular frequency $\omega$.

The boundary conditions near BHs are 

\begin{align}
    \bar \beta ^{\bar i}|_{S_{\rm BH}} = &\,\, \bar \alpha \bar \Psi^{-2} \tilde s^{\bar i} + \omega \xi^{\bar i}_{\rm BH},\\
    \tilde s^{\bar i} \tilde D_{\bar i} (\bar \alpha \bar \Psi)|_{S_{\rm BH}} = &\,\, 0,\\
    \tilde s ^{\bar i} \tilde D_{\bar i} \bar \Psi |_{S_{\rm BH}} = &\,\, - \frac{\bar \Psi }{4} \tilde D^{\bar i} \tilde s_{\bar i} - \frac 14 \bar \Psi^{-3} \hat A_{\bar i \bar j} \tilde s^{\bar i} \tilde s ^{\bar j},
\end{align}
where $\xi_{\rm BH}$ is vector field along the azimuthal direction with reference to BH center and $\tilde s$ is unit normal vector on surface $S_{\rm BH}$.

Once the BBH spacetime is solved on $(\bar t,\bar x,\bar y,\bar z)$ coordinate, we then obtain the metric in the simulation frame $(t,x,y,z)$ by Eq. \ref{eq_metric_trans_op}-\ref{eq_metric_trans_ed}. {Since the binary spacetime is constructed using a helliptical Killing vector (i.e, fixed orbit in the absence of gravitational wave emission), we can efficiently rotate the spacetime on a fixed circular orbit. We do so by explicitly Lie-dragging the spacetime metric during the simulation, using the comoving shift vector ${\boldsymbol{\beta}} = {\boldsymbol \omega} \times {\boldsymbol r}$.} 

\subsection{Evolution of scalar field}
\label{sec:formulation_scalar}
The {massive} complex scalar field {(mass $m_\phi = \hbar \mu$)} is evolved both in the simulation frame and the BBH frame. It satisfies the massive Klein-Gordon equation $(\Box - \mu^2)\phi = 0$ on the given background metric. The conjugate momentum, projection of the field gradient along $n^\mu$, $\Pi := n^\mu \nabla_\mu \phi$ is introduced to reduce the system to first-order. The system of equations for the scalar field is \cite{Okounkova:2017yby}:
\begin{align}
    \label{eq_scalar_evolution}
    \partial_t \phi =& \,\, \alpha \Pi + \beta^i \partial_i \phi,\\
    \partial_t \Pi = & \,\, \alpha \partial^2 \phi + \alpha (K\Pi - \gamma^{ij} \Gamma^k_{ij} \partial_k \phi - \mu^2 \phi) + \partial_i \phi \partial^i \alpha + \beta^i \partial_i \Pi,
\end{align}
where $K = \gamma^{ij}K_{ij}$ is the trace of the extrinsic curvature and $\Gamma^k_{ij}$ are the Christoffel symbols corresponding to the spatial metric $\gamma_{ij}$.

We performed simulations both for coordinate $(t,x,y,z)$ in the rest frame of the scalar field background (while comoving with the boosted BBH system) and the “barred” coordinate $(\bar t, \bar x, \bar y, \bar z)$ in the rest frame of the BBH. The initial and boundary conditions for the two frames are different: In the rest frame of the scalar field, we impose a uniform initial condition $\phi|_{t=0} = \phi_0,\, \partial_t \phi |_{t=0} = i\mu \phi_0 $ and an oscillatory boundary condition $\phi|_{r=r_{\rm boundary}} = \phi_0 e^{i\mu t}$. In the rest frame of the BBH, we impose a propagating wave for initial and boundary conditions $\phi|_{t=0} = \phi_0 e^{-ikx},\, \partial_t \phi |_{t=0} = i\mu \phi_0 e^{-ikx},\, \phi|_{r=r_{\rm boundary}} = \phi_0 e^{i\mu t-ikx} $. The rest frame energy density of the field in its comoving frame is $\rho_{\rm asymp} = \frac{1}{2}\mu^2|\phi_0|^2 (1+k^2)$ at infinity, which defines the asymptotic “cloud” density in our setup. We emphasize that the spacetime metric is held fixed to the BBH solution by XTCS method (with appropriate boost and advection), so that the scalar fields stress-energy does not source Einstein's equations. This test-field approximation is valid to first order in the scalar field strength: metric perturbation induced by the field would enter at $\mathcal{O}(|\phi|^2)$ and the secondary effect of that on the scalar field evolution itself is neglected in our simulations.

\begin{figure}
    \centering
    \includegraphics[width=0.9\linewidth]{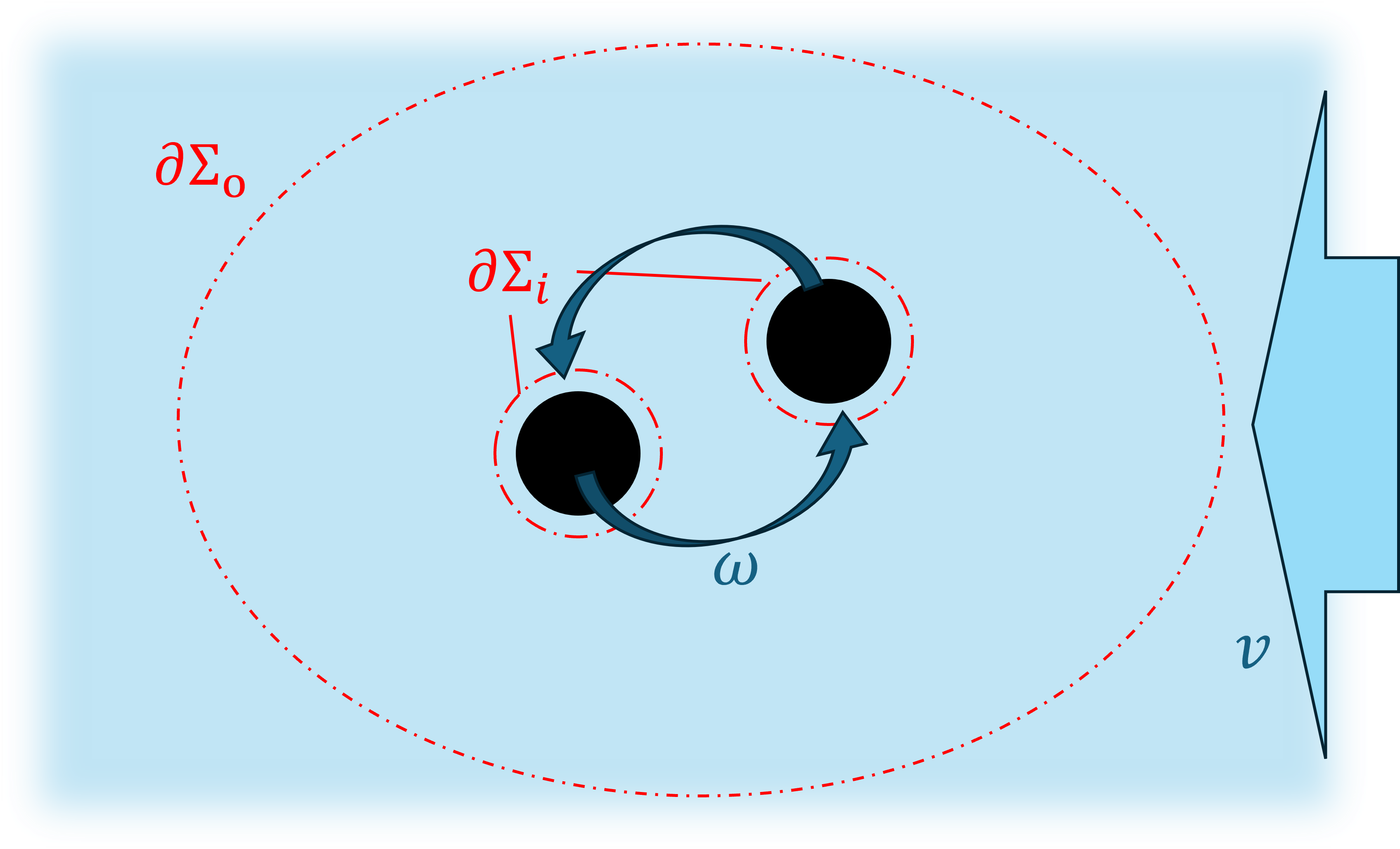}
    \caption{Illustration of our numerical setup. Two black holes orbit each other with angular frequency $\omega$. The red dashed curves denote the inner excision surfaces $\partial\Sigma_i$ that enclose each horizon and the outer integration surface $\partial\Sigma_o$ that bounds the control volume $\Sigma_o-\Sigma_i$ used for the volume and surface integrals. The light-blue region depicts the homogeneous scalar background flowing from right to left with asymptotic velocity $v$.}
    \label{fig:illustration}
\end{figure}

\subsection{Diagnostics}
{
In order to evaluate the impact of the scalar field onto the binary system, we need to introduce a set of diagnostic tools designed to measure drag forces and torques exterted onto the system.\\
Our starting point is the energy momentum tensor of the scalar field,
\begin{equation}
    T_{\mu\nu} = \frac 12 \left( \nabla_\mu \phi \nabla_\nu \phi^*  + \nabla_\mu \phi^* \nabla_\nu \phi -g_{\mu\nu} \nabla^\lambda \phi \nabla_\lambda \phi^*  \right)\,,
\end{equation}
which is conserved according to
\begin{align}
    \nabla_\mu T^{\mu\nu}=0\,.
\end{align}
In numerical implementations we need to rewrite the integrands in $3+1$ ADM quantities. We define the usual decomposition of the energy momentum tensor:

\begin{equation}
    T_{\mu\nu} = S_{\mu\nu} + 2 S_{[\mu} n_{\nu]} + \rho n_\mu n_\nu,
\end{equation}
where the 3-stress, momentum, and energy densities are
\begin{align}
    S_{ij} =&\, \gamma_{i\mu} \gamma_{j\nu} T^{\mu\nu},\\
    S_{i} =&\, -\gamma_{i\mu} n_{\nu}T^{\mu\nu},\\
    \rho = & \, n_\mu n_\nu T^{\mu\nu}.
\end{align}

Energy, angular momentum and linear momentum can then be defined using an approximate Killing vector, $\zeta^\mu$, which can then implies that for currents,
\begin{align}
    \mathcal{J}^\mu = T^{\mu\nu}\zeta_\nu\,,
\end{align}
the following modified conservation law should hold
\begin{align}\label{eqn:Killing}
    \nabla_\mu \mathcal{J}^\mu = T^{\mu\nu}\nabla_\mu \zeta_\nu\,.
\end{align}
Since the interaction of the scalar field with the binary does not have or conserve any symmetry, the right hand side of Eq. \eqref{eqn:Killing} does not vanish, since $\zeta$ is not a real Killing vector.

By integrating this equation over the spatial domain, $\Sigma$, with surface normal, $\hat s^i$, it can be interpreted as a continuity equation for a charge, 
\begin{equation}
\label{eq_J_conserve}
\partial_t \left(\int_\Sigma \sqrt{\gamma} \mathcal{Q} d^3 x  \right) = -\int_{\partial \Sigma} \mathcal{F} dA + \int_\Sigma \sqrt{\gamma} \mathcal{S} d^3x ,
\end{equation}
where the charge density, flux and source are
\begin{equation}
    \mathcal{Q} = -n_\mu \mathcal{J}^\mu,
\end{equation}
\begin{equation}
    \mathcal{F} = \alpha \hat s_i \mathcal{J}^i ,
\end{equation}
\begin{equation}
\mathcal{S} = \alpha T^{\mu\nu}\nabla_\mu \zeta_\nu . 
\end{equation}  
}

When computing these integrals, we have to account for the fact that our black holes have excision boundary conditions, and that we are using a finite size simulation domain. As a consequence, the integral is instead over a finite ball $\Sigma_o$ with a cutoff radius $r^{\rm out}$. The singularities are removed by subtracting $\Sigma_i$ with the boundary $\partial \Sigma_i$ being two spheres of radii $r^{\rm in}_1,r^{\rm in}_2$ covering the BBH horizons. Following Gauss's law, we should account for finite boundary effects by including the flux across $\partial \Sigma_{o,i}$, i.e. a surface integral of the flux into the unit normal of surface $\vec J\cdot \hat n$ over the boundaries. The integral over the outer boundary gives the net momentum input to the system and the integral over the inner boundaries gives the accretion effect.

{In the following sections, we will explicitly compute this source term for the different properties (linear momentum, angular momentum and energy) of the binary.}

\subsubsection{Gravitational momentum drag}

{
To compute the change in linear momentum of the binary, i.e., the drag exerted by the scalar field, we choose $\zeta_i^\nu = \delta^\nu_i$ corresponding to the unit vector in the $x^i$ direction.

We then find that the linear momentum, {$P_i$},
\begin{align}    
P_i = \int_\Sigma \sqrt{\gamma} \mathcal{Q}_i d^3 x\,,
\end{align}
changes according to
\begin{align}
\label{eq_dPidt}
\partial_t P_i =
     \underbrace{\int_{\Sigma_o - \Sigma_i} \,T^{\mu}_\nu\,\nabla_\mu(\delta^{\nu}_{\;i})\,\sqrt{-g}\,d^3x}_{F_i^{\rm drag}} \;+\; \underbrace{\int_{\partial\Sigma_i} \alpha\,T^{j}_{i}\,dA_j}_{F_i^{\rm accretion}}~. 
\end{align}
Here, we have introduced two terms, namely the drag force, $F_i^{\rm drag}$, and the accretion force, $F_i^{\rm accretion}$, coming from the source $\mathcal S$ and flux terms $\mathcal F$, respectively. 

In terms of the ADM variables, these can be re-expressed as}
\begin{equation}
\label{eq_drag}
    F^{\rm drag}_i = \int_{\Sigma_o - \Sigma_i} \hspace{-0.6cm} d^3x \sqrt{\gamma} \, \left( -\rho \partial_i \alpha + S_j \partial_i \beta^j + \alpha S^k_j \Gamma^j_{ki} \right),
\end{equation}
\begin{equation}
    F^{\rm accretion}_i = \int_{\partial \Sigma_i} dA\, \hat s_j \left( \alpha \gamma^{jk} S_{ki} - \beta^j S_i \right),
\end{equation}
where $\Gamma^{j}_{ki}$ is the Christoffel symbol corresponding to the spatial metric $\gamma_{ij}$, $A$ is surface area element, and $\hat s$ is unit normal vector on the surface. Note that $\partial \Sigma_i$ includes two spheres and normal vectors are defined with respect to each sphere.

For diagnostics, we also extract the surface integral over the outer boundary, which gives the net momentum flux into the system
\begin{equation}
    F^{\rm out}_i = \int_{\partial \Sigma_o} dA\, \hat s_j \left( \alpha \gamma^{jk} S_{k i} - \beta^j S_i \right).
\end{equation}
The total effect of $F^{\rm drag},F^{\rm accretion},F^{\rm out}$ is responsible for the momentum change of scalar field in the domain
\begin{equation}
    \Delta P_i = \int_{\Sigma_o - \Sigma_i} S_i \sqrt{\gamma} \,\,d^3x.
\end{equation}
Therefore we should expect the balance between momentum change and the three forces. Details are given in the Appendix \ref{app:conservation}.

\subsection{Angular momentum flux and dragging torque}
{To compute torques acting on the binary, we compute the net angular momentum flux onto the binary. To this end, we define $\zeta^\nu =\zeta_\phi^\nu = \left(0,-(y-y_{\rm com}), x-x_{\rm com},0\right)$, where $(x_{\rm com}, y_{\rm com},0)$ is the center of mass of the binary.}  The resulting orbital angular momentum is then given by
\begin{equation}
    \mathcal{L}_i = \int_{\Sigma_o - \Sigma_i} \epsilon_{ijk} x^j S^k \sqrt{\gamma} d^3x\,,
\end{equation}
the flux $\mathcal{F}$ as the torque due to the accretion of angular momentum through the BBH surface $\partial \Sigma_i$ and outer boundary $\partial \Sigma_o$ 
\begin{equation}
    \tau^{\rm accretion}_i = \int_{\partial \Sigma_i} dA\, \hat s_j x^l \left( \alpha \gamma^{jk} S_{k}^m - \beta^j S^m \right) \epsilon_{ilm},
\end{equation}
\begin{equation}
    \tau^{\rm out}_i = \int_{\partial \Sigma_o} dA\, \hat s_j x^l \left( \alpha \gamma^{jk} S_{k}^m - \beta^j S^m \right) \epsilon_{ilm},
\end{equation}
and the source $\mathcal{S}$ as the torque due to drag distributed over the bulk of space
\begin{equation}
    \tau^{\rm drag}_i = \int_{\Sigma_o - \Sigma_i} \hspace{-0.6cm} d^3x \sqrt{\gamma} \, x^l\left( -\rho \partial^m \alpha + S_j \partial^m \beta^j + \alpha S^k_j {\Gamma^j_{k}}^m \right) \epsilon_{ilm}.
\end{equation}
Similar to the momentum flux conservation, the angular momentum obeys a conservation law following Eq. \ref{eq_J_conserve}
\begin{equation}
    \label{eq_L_conserve}
    \partial_t \mathcal L_i = \tau^{\rm out}_i - \tau^{\rm drag}_i - \tau^{\rm accretion}_i,
\end{equation}
which states that the outer boundary provides some inflow of angular momentum by $\tau^{\rm out}$, after the drag dissipation to spacetime by $\tau^{\rm drag}$ and accretion to BBH by $\tau^{\rm accretion}$, the residual contribute to the total angular momentum change in the scalar field $\partial_t L_i$. The $\tau^{\rm drag}+\tau^{\rm accretion}$ is the total back reaction to spacetime.

\subsection{Mass and charge accretion}

{Finally, we can also quantify the amount of accreted scalar field energy onto the binary.}
Here, we assume that $\zeta_\mu = n_\mu$, so that $\mathcal{-Q} =\rho = n_\mu n_\nu T^{\mu\nu}$.
Thus, following the form of Eq. \ref{eq_J_conserve} (also demonstrated in \cite{clough2021continuity}), we have the energy flux
\begin{equation}
    \mathcal{F}^{E} = \hat s_i (\alpha S^i - \rho \beta^i),
\end{equation}
and dissipation
\begin{equation}
\begin{aligned}
    \mathcal{S}^E = & \,\, \alpha K^{ij}S_{ij} + S^i(\alpha \Gamma^{j}_{ij} - \partial_i \alpha)\\
    & + \rho \left( \partial_i \beta^i - \alpha K - \frac{1}{\alpha} \beta^i \beta^j K_{ij} \right).
\end{aligned}
\end{equation}

Due to the symmetry of the scalar field under $\phi \rightarrow \phi e^{i\theta}$ transformation, the field endows a conserved scalar charge. The Noether current corresponding to the charge is 
\begin{equation}
    j_\mu = \phi \nabla_\mu \phi^* - \phi^* \nabla_\mu \phi .
\end{equation}
The current conservation of this complex scalar field charge is exact, $\nabla_\mu j^\mu = 0$, everywhere in the domain. Therefore there's no dissipation term following Eq. \ref{eq_J_conserve} for the complex scalar charge \cite{croft2023local},
\begin{equation}
    -i \mathcal Q^{\phi} = \phi^* \Pi - \phi \Pi^*,
\end{equation}
with conserved flux across surfaces
\begin{equation}
    -i\mathcal{F}^\phi = \alpha \hat n_i(\phi D^i \phi^* - \phi^* D^i \phi) - \hat n_i \beta^i \mathcal{Q}^\phi .
\end{equation}

This allows us to obtain the mass $\dot{M}$ and scalar charge $\dot{Q}$ accretion onto the BBH as
\begin{equation}
    \dot M^{\rm acc} = \int_{\partial \Sigma_i} dA \hat s_i (\alpha S^i - \rho \beta^i) + \int_{\Sigma_o - \Sigma_i} d^3 x \sqrt{\gamma} \mathcal{S}^E\,,
\end{equation}
\begin{equation}
    \dot Q^{\rm acc} = i\int_{\partial \Sigma_i} dA \hat s_i \left( \alpha\phi D^i \phi^* - \alpha \phi^* D^i \phi -\beta^i \phi^* \Pi +\beta^i \phi \Pi^* \right)\,.
\end{equation}

\begin{figure*}
    \centering
    \includegraphics[width=0.9\linewidth]{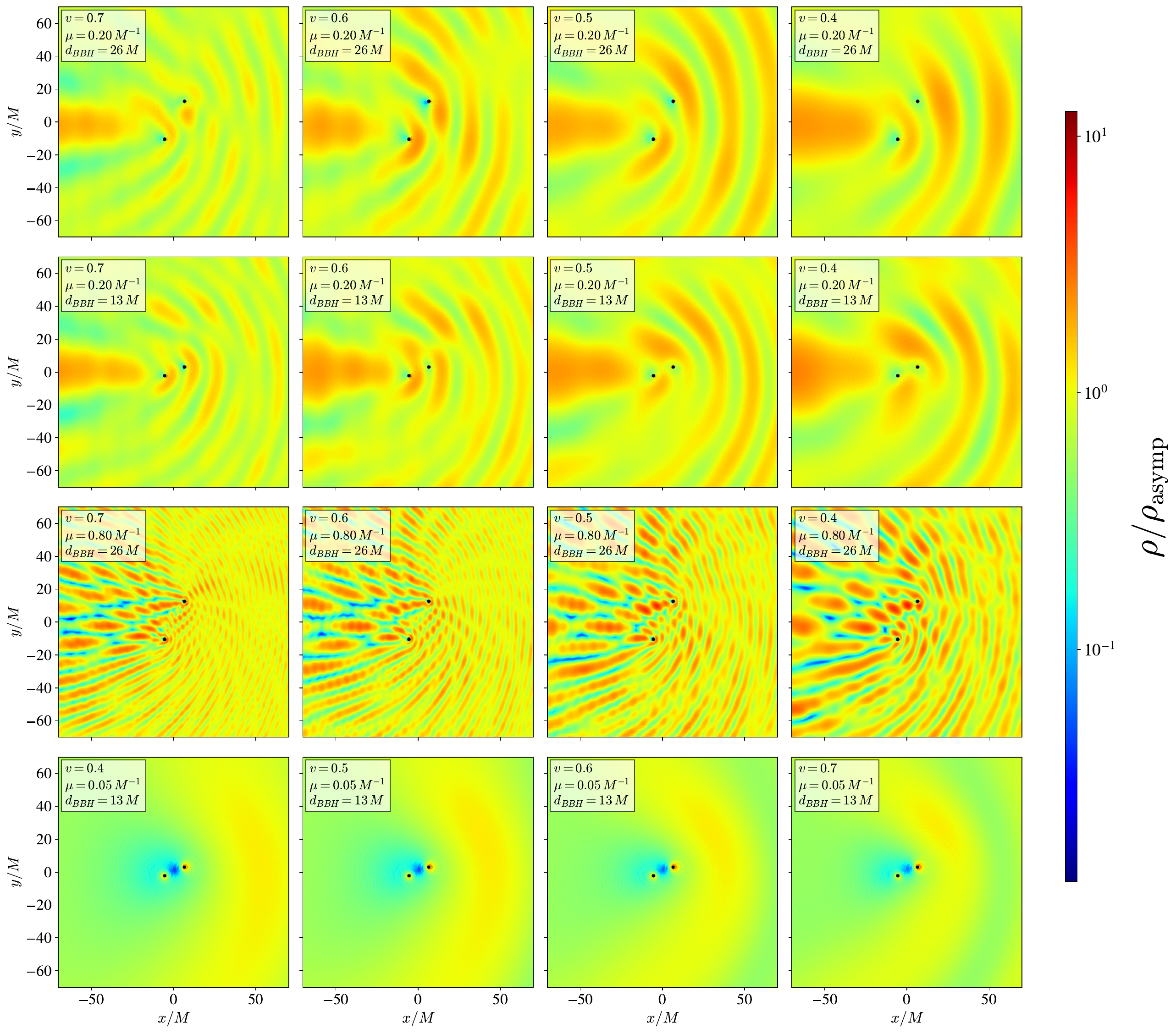}
    \caption{Scalar field energy density, $\rho$, the orbital plane for a quasi-stationary state (at time $t=4500\,M$) for different background velocities $v$, scalar field masses $m_\phi=\hbar \mu$, and binary separations $d_{\rm BBH}$. Columns (left to right) correspond to $v=0.7,\,0.6,\,0.5$, and $0.4$. Rows (top to bottom) show the four parameter regimes: (1) $\mu M = 0.20,\; d_{\rm BBH}=26M$; (2) $\mu M = 0.20,\; d_{\rm BBH}=13M$; (3) $\mu M = 0.80,\; d_{\rm BBH}=26M$; (4) $\mu M = 0.05,\; d_{\rm BBH}=26M$. Black circles mark the black hole horizons and the color bar indicates $\rho$ on a logarithmic scale. }
    \label{fig:rho_energy_velocity_comparison}
\end{figure*}

\begin{figure}
    \centering
    \includegraphics[width=\linewidth]{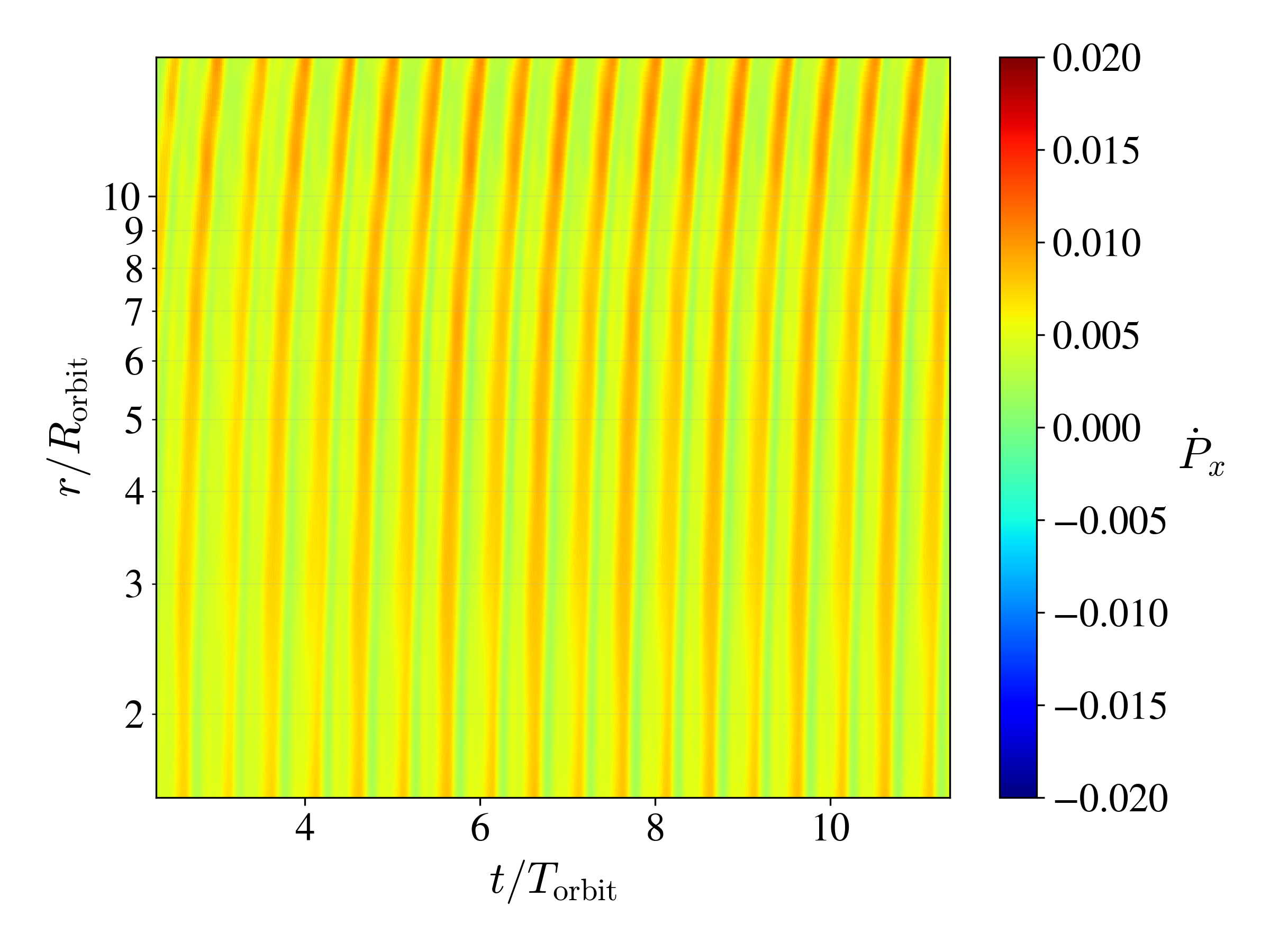}
    \includegraphics[width=\linewidth]{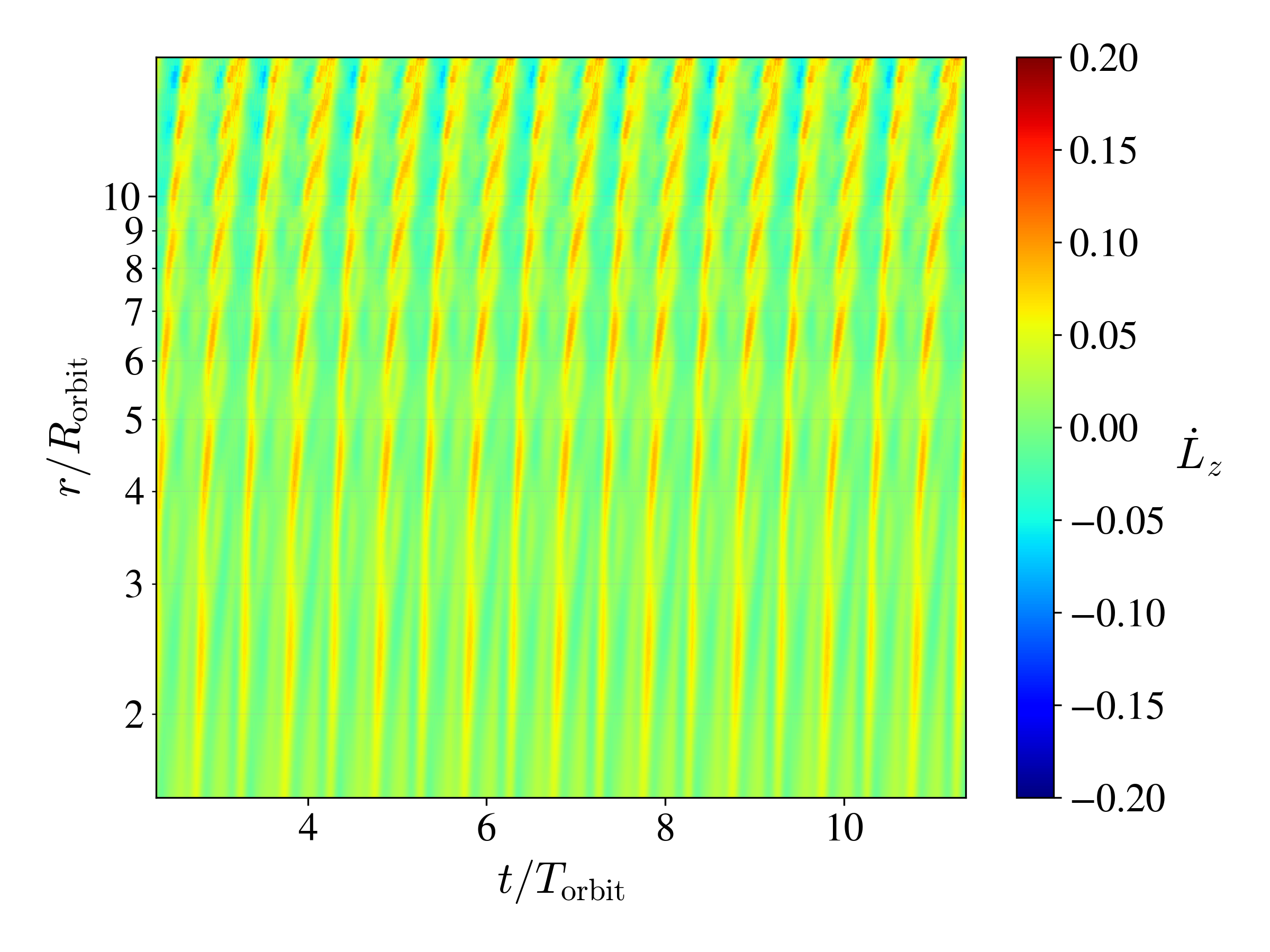}
    \caption{Evolution of momentum flux $\dot P_x$ (upper panel) and angular momentum flux $\dot L_z$ (lower panel) distribution for the run with $v=0.5, \mu M = 0.2, d_{\rm BBH} = 26M$ (where $R_{\rm orbit} = d_{\rm BBH}/2 = 13 M$). The temporal periodicity is same as the orbital period.}
    \label{fig:heatmap}
\end{figure}



\subsection{Numerical Methods}
\label{sec:method}

Our simulations evolve a complex scalar field on a fixed spacetime metric representing a binary black hole. In this section, we describe the numerical setup in detail, including the numerical construction of the binary black hole initial data and the evolution scheme for the scalar field. We also summarize the codes and algorithms employed, and we outline the validation tests performed to ensure the reliability of our results.

To model the spacetime of BBH, we utilize a spectral elliptic solver to construct constraint-satisfying initial data for two black holes in quasi-circular orbit using the XCTS formalism \cite{Caudill:2006hw}. In particular, we make use of the \textsc{FUKA} (Frankfurt/\textsc{Kadath}) initial data framework \cite{Papenfort:2021} to generate the binary black hole metric. \textsc{FUKA} employs spectral methods built on the \textsc{Kadath} library \cite{Grandclement:2010} to solve the Einstein constraint equations for arbitrary mass ratios, spins, and orbital parameters. We generated initial data for an unequal-mass binary with mass ratio $ q = M_1 / M_2 = 0.8$. $M=M_1+M_2$ is the total mass of the binary and also the geometrized unit of the simulation variables in our following discussions. The binary is placed on a quasi-circular orbit. We tested different initial separations chosen such that we cover the parameter range where the binary is comparable/much larger/much smaller than the scalar field de Broglie wavelength, ${\lambda}$.
Because we do not evolve the spacetime dynamically (the metric is held fixed during the scalar field evolution), the binary black hole orbit does not shrink due to gravitational-wave emission in our simulations. Effectively, the BHs are treated as if on a fixed background trajectory for the duration of the run. This approximation is valid for our purposes, since we focus on the instantaneous forces exerted by the scalar field on the BHs and the resulting energy and momentum exchange. We choose coordinates co-rotating with the binary for setting up the initial data, but then transform to an inertial frame for the evolution.
\begin{figure}[t]
    \centering
    \includegraphics[width=\linewidth]{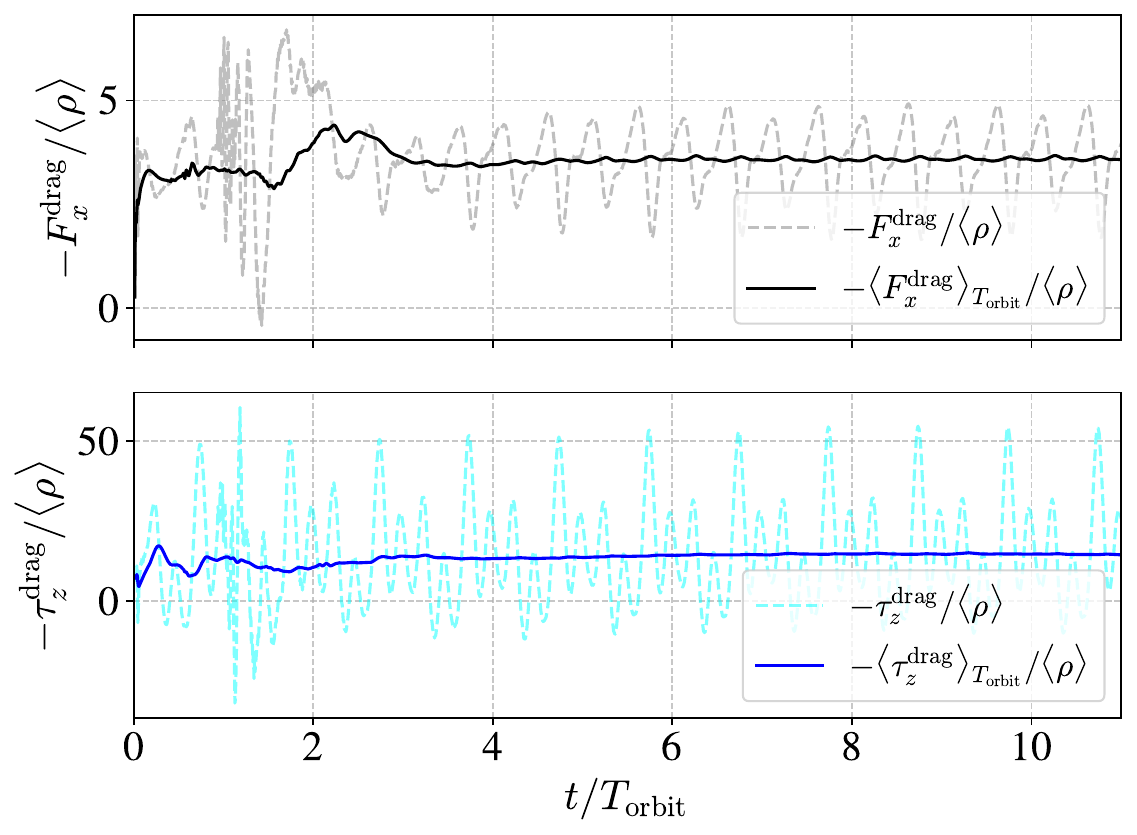}
    \caption{Orbital averaging of the drag force $F_x$ and torque $\tau_z$ for the run with $v=0.5, \mu M = 0.2, d_{\rm BBH} = 26M$.}
    \label{fig:drag_averaging}
\end{figure}

We evolve the complex scalar field on the fixed orbit BBH spacetime using the same code infrastructure previously employed in Ref. \cite{MostPhilippov:2022,Most:2023unc} and its GPU-based version \cite{Most:2024eig}, which is built on the block-structured adaptive mesh refinement framework \textsc{AMReX} \cite{Zhang:2019}. We have extended this code to also solve the Klein-Gordon equation on the numerical background metric provided by \textsc{FUKA}. Specifically, Eq. \ref{eq_scalar_evolution} is evolved using centered finite-difference techniques with fourth-order accuracy in time and space. We employ an explicit fourth order Runge-Kutta time integrator and centered spatial differencing on a Cartesian grid. The grid is refined using static mesh refinement centered around each BH and in regions of steep scalar gradients, ensuring adequate resolution of the scalar field near the BH horizons and in the wake regions. 
The numerical implementation is fully GPU-enabled. {While the accretion process on the black hole binary system is naturally three-dimensional, we follow similar works for hydrodynamic torque calculations in circumbinary \cite{munoz2020circumbinary} and AGN disk \cite{li2023hydrodynamical}, which have calculated drag forces and torques using using vertically integrated accretion flows. This allows us to provide a systematic coverage of the large relevant parameter space, while simulating sufficiently many orbits to obtain converged results in this initial exploration.}

{The numerical resolution and outer boundary prescription of our simulation is chosen to yield convergent results.} First, we performed a resolution convergence test for a reference simulation ($\mu M=0.2$, $d_{\rm BBH}=26M$, $v=0.5c$). By repeating this simulation with successively finer grids and comparing the results, we verified convergence of the evolved scalar field and of the computed force on the BHs. Details of the convergence test are presented in Appendix~\ref{app:convergence}. Next, we verify the boundary effect at the inner extraction surface. The individual contribution to the force from volume integral and surface integral varies with different inner extraction radius $r_{\rm in}$ while the total force is independent of $r_{\rm in}$, as detailed in Appendix~\ref{app:boundary}. Finally, we verify the conservation of linear and angular momentum in the simulation domain, as explained in Appendix~\ref{app:conservation}.

We conducted four series of simulations to explore the parameter space of scalar field mass $\mu M$, binary separation $d_{\rm BBH}$, orbital frequency $\omega M$, and background velocity $v$. Table~\ref{tab:sim_setup} summarizes the key parameters for each simulation series. The computational domain is a cubic box with side length $L_{\rm box}$, centered on the binary's center of mass. We use $N^2$ grid points for the base grid, with additional refinement levels near the black holes. The de Broglie wavelength of the scalar field $\lambda = 2\pi/k$, where $k = \gamma v \mu$ is the wavenumber and $\gamma = 1/\sqrt{1-v^2}$ is the Lorentz factor in the simulation frame, depends on both the scalar mass and the background velocity. For each series, we performed simulations with background velocities $0.2c < v < 0.7c$ to study the velocity dependence of the drag force and torque.

\begin{table}[htbp]
\centering
\begin{tabular}{cccccccc}
\hline
& $\mu M$ & $d_{\rm BBH}/M$ & $\omega M$ & $N$ & $L_{\rm box}/M$ & $\lambda/M$ & refinement levels\\
\hline
Case A & 0.2 & 26.0 & 0.007 & $256$ & 1600 & $31.4/(\gamma v)$ & 6 \\
Case B & 0.2 & 13.0 &0.021 & $256$ & 1600 & $31.4/(\gamma v)$ & 6 \\
Case C & 0.8 & 26.0 & 0.007 & $320$ & 800 & $7.8/(\gamma v)$ & 5 \\
Case D & 0.05 & 13.0 &0.021 & $416$ & 6400 & $125.7/(\gamma v)$ & 8 \\
\hline
\end{tabular}
\label{tab:sim_setup}
\end{table}

\begin{figure*}
    \centering
    \includegraphics[width=0.9\linewidth]{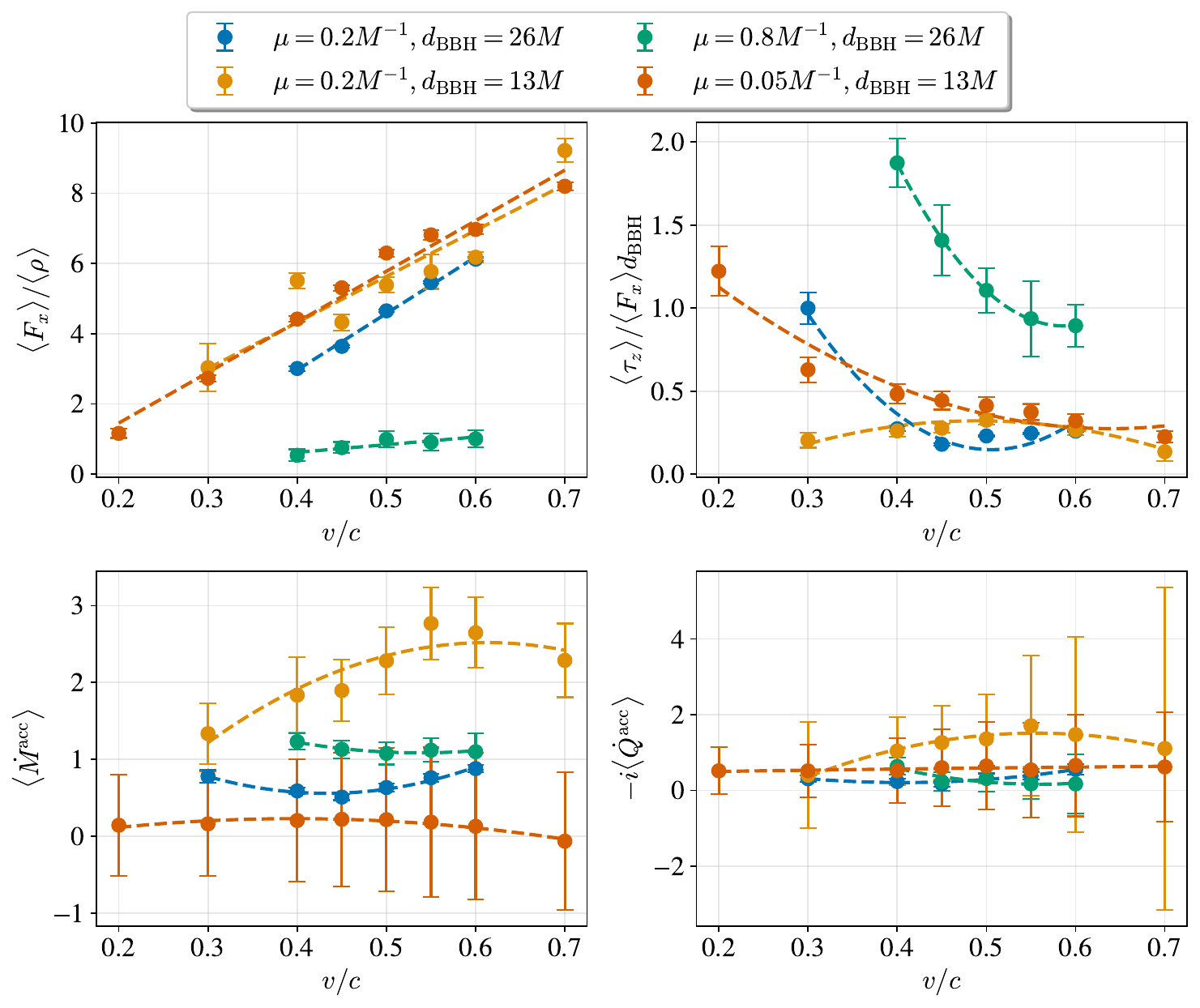}
    \caption{{Overall scalar drag properties. }Velocity dependence of friction force $F_x$, torque $\tau_z$, mass accretion $\dot{M}^{\rm acc}$ and charge accretion $\dot{Q}^{\rm acc}$ for the four parameter regimes studied in this work.}
    \label{fig:combined_quantities_vs_velocity}
\end{figure*}

\section{Results}
\label{sec:results}

In this section, we present the main results of our simulations, focusing on the structure of the scalar field distribution around the binary and the forces and torques experienced by the black holes. Spatial and temporal periodicity of momentum and angular momentum flux are analyzed. We also estimate the mass and charge accretion onto the black holes, as well as the dependence on the binary and scalar field parameters. 

{Before diving into quantitative estimates, we briefly clarify the global behavior of the system.
{We point out that for the relativistic speeds we use, Lorentz contraction is, in fact, important, though we will extrapolate also to the mildly relativistic limit later.}
On scales of the orbital separation, the incoming scalar field wind is almost a plane wave. This scenario is similar to a (rotating) diffraction grating in optics. Consequently, we  can identify three limiting cases depending on the ratio, $\lambda / d_{\rm BBH}$, of the de Broglie wavelength to the orbital separation, where we assume that the wavelength of the field is comparable to the black hole size, $r_{\rm BH}$. For  $\lambda \ll d_{\rm BBH}$, we are in the Fraunhofer limit, where diffraction of the field will occur. For $\lambda \simeq d_{\rm BBH} \simeq r_{\rm BH}$, we are in the Fresnel limit, where we expect the strongest amount of nonlinearities to appear. 
{In this limit, our results appear consistent with the single BH case of \cite{Wang:2024cej,Traykova:2021} in terms of wake formation.}
For $\lambda \gg d$, no diffraction will occur. In Fig. \ref{fig:rho_energy_velocity_comparison}, we present these limiting cases. For scalar field masses $\mu M =0.05$, the wavelength of the scalar field is much larger than the horizon size and the orbital separation, $d$. In this regime, the scalar field propagates through the binary unimpeded.}


{Focusing on a single case,} we extract the momentum and angular momentum flux across a series of surface radii at each time snapshot for a representative case ($v=0.5$, $\mu M=0.2$, $d_{\rm BBH}=26M$). In the spatial-temporal distribution of the fluxes Fig. \ref{fig:heatmap}, we can clearly identify the periodicity of $T_{\rm orbit} = 2\pi /\omega \approx 870M $ both in the momentum flux $\dot P_x$ and angular momentum flux $\dot L_z$. We can see a surge of flux appearing every $T_{\rm orbit}/2$, and alternate between a slightly brighter and a slightly fainter one, which is consistent with our mass ratio $q=0.8$ close to equal mass. Such periodicity extends in space up to $>15 R_{\rm orbit}$ close to the outer boundary of our integration domain, showing that the BBH physical effect dominates in our simulation domain. In the radial direction, we can also see the angular momentum flux oscillates approximately at a period of $d_{\rm BBH}$, showing a radiation pattern modulated by the binary separation.


Similar periodicity is present for volume integrals of the drag force. By averaging the fluxes, $F_x$, over an orbital period, we extract nearly constant values for the drag force and torque, $\tau_z$. Figure~\ref{fig:drag_averaging} shows {these quantities as a time series. After an initial transient (around $t\sim4000\,M$ in this example), the system settles into a steady state}.{ We can again see the dual beat frequency in the drag force and torque, corresponding to the mass asymmetry of the binary. The orbital average of both quantities is remarkably constant in time, indicating a converged state. This is consistent with the fact that our simulations span multiple Bondi times $\tau_{\rm Bondi} = 2 M/v^3$.}





The drag force, torque, and mass/charge accretion rate are sensitive to the velocity of the incoming DM wind. {We, therefore, provide a full parameter survey of the drag force and torque properties in Fig. \ref{fig:combined_quantities_vs_velocity} for models considered in this work.} 
For a representative case ($\mu M=0.2$, $d_{\rm BBH}=26M$), the orbit-averaged force over time is shown in Fig. \ref{fig:combined_quantities_vs_velocity} for a series of velocity $0.3 < v/c < 0.7$. Near $v\sim 0.5$, the change of wind velocity corresponds to approximately linear change of steady state force, as fitted in the blue dashed line in top left panel of Fig. \ref{fig:combined_quantities_vs_velocity}. Below $v=0.3c$, the simulation suffers from more boundary effects, as the effective Bondi radius becomes comparable to our domain size, and may not be accurate.
Above $v=0.7c$ the force appears to saturate. {We provide a more detailed assessment of the convergence of these results in Appendix \ref{app:convergence}.} 
{This trend generically persists also when varying the incoming wavelength, however the drag force is suppressed in the strongly interference dominated Fraunhofer like regime ($\mu =0.8/M$).}

We have similarly extracted the drag torque to the binary. We see a larger error bar compared with the drag force (cf. top left and right panels of Fig. \ref{fig:combined_quantities_vs_velocity}). Since the torque scales as $\vec \tau \sim \vec r \times \vec F \propto r$, the fluctuations are magnified as we go to larger radii. Due to such effect, torque are usually only extracted up to $r=40M$, as \cite{croft2023local}. For our simulation, we confirmed the numerical accuracy by monitoring the angular momentum conservation laws (see Appendix \ref{app:boundary}). The conservation is well preserved at $r=45M$ and have small deviations up to $r=650M$.

The torque clearly depends nonlinearly on velocity, with a minimum or maximum reached near $0.4c \sim 0.5c$. For the two simulations using larger separation binaries, the minimum is reached near $v=0.5c$ while for smaller separation binaries, the maximum is near $v=0.5c$. {At all times the net torque acts on the system so as to spin it down. It will remain interesting to confirm how these results extend to mildly and non-relativistic velocities.}

\begin{figure}
    \centering
    \includegraphics[width=\linewidth]{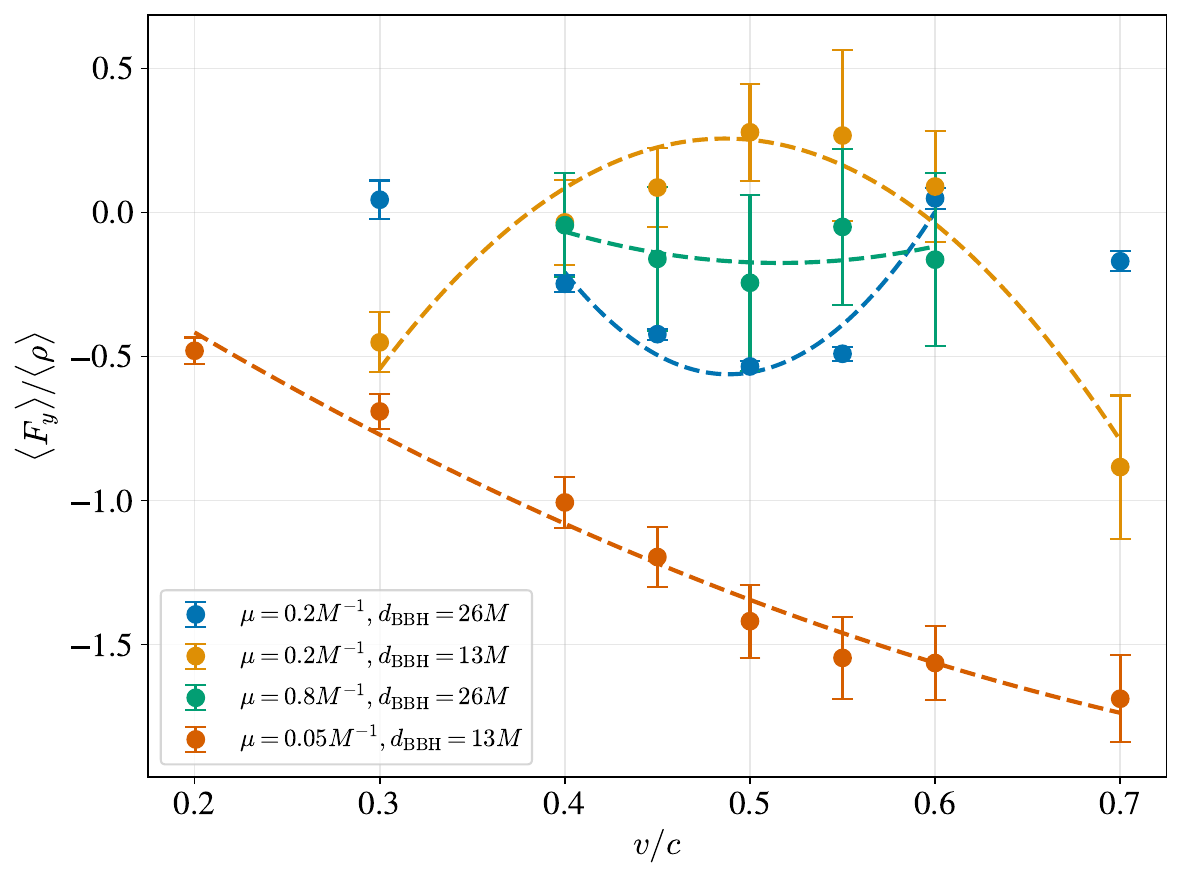} 
    \caption{ Magnus force, $F_y$, transverse to the wind and spinning directions exerted on the binary system. }
    \label{fig:Fy_vs_vel}
\end{figure}

{Spinning bodies moving in viscous matter experiences a Magnus effect, a force perpendicular to both moving and spinning directions. The gravitational analogue for scalar field matter was recently studied for Kerr black holes \cite{Wang:2024cej,Dyson:2024qrq}, and for magnetohydrodynamical accretion in Ref. \cite{Kim:2024zjb}. In our case, the black hole binary moves in +x direction while spinning in +z direction. We found $F_y < 0$ for $0.3<v<0.6$ as shown in Fig. \ref{fig:Fy_vs_vel}. The sign is the same as the case for Kerr black holes, i.e. in the ``anti-Magnus'' direction, opposite from hydrodynamical cases \cite{Kim:2024zjb}. However, we find (within considerable error bars) that for some parameter ranges the sign of the Magnus force for the binary can change. 
The absolute value depends nonlinearly on velocity and reaches maximum near $v=0.5$, consistent with the findings in \cite{Wang:2024cej}.  }

The black holes also steadily accrete scalar field. We measure a constant mass accretion rate $\dot{M}^{\rm acc}$ (Fig. \ref{fig:combined_quantities_vs_velocity}). T
he accretion is generally small, but becomes noticeable for heavy scalars at low velocity. For example, for $\alpha_s=1$ and $v=0.2$ we infer an accretion efficiency $\lambda\sim0.5$, which helps explain the total force in that regime (consistent with analytical expectations \cite{Traykova:2023letter}). In the opposite regime (light scalar or $v\gtrsim0.5$), the accretion contribution is negligible ($\lambda\approx0$). We also verify that the scalar Noether charge accretion $\dot{Q}^{\rm acc}$ satisfies $\dot{Q}^{\rm acc}\approx\dot{M}^{\rm acc}/\mu$ and is always positive, confirming that the BHs absorb (and do not emit) scalar particles.

\section{Conclusion}
\label{sec:discussion}

We have quantified the drag force on a BBH moving through a relativistic scalar dark matter medium. Our simulations show that the drag can be described by a modified dynamical friction law that includes relativistic and accretion effects. These results generalize previous single-BH analyses \cite{Traykova:2021,Traykova:2023letter,Wang:2024cej} to the case of a relativistic binary, capturing the combined effect of the two wakes and their interference.

Astrophysically, the drag forces we consider are extremely small under typical dark matter densities. {Taking an analytic approximation \cite{Traykova:2021}, the drag force $F \sim 4\pi \rho (\frac Mv)^2$, while the energy dissipation from GW radiation is $\dot E_{\rm GW} \sim \frac{32}{5} \frac{\mu^2}{(1+\mu)^4} ( r_{\rm BBH})^{-5}$ \cite{boyle2008high}}. For example, with $\rho\sim10^{-20}\,$g/cm$^3$ $\sim 10^{-38}M^{-2}_{\odot}$, the energy dissipation from gravitational friction is orders of magnitude below that from gravitational-wave radiation for a stellar mass binary $M\sim 10M_{\odot}$ in the separation we considered $\omega r_{\rm BBH} \sim 0.1$. However, in scenarios with very dense scalar clouds (e.g.\ a dark matter spike around a massive BH) \cite{hannuksela2019probing}, the effect could become more significant, possibly detectable by future detectors like LISA \cite{hannuksela2019probing}. Thus our findings provide quantitative benchmarks for how ultralight dark matter might imprint subtle signatures on gravitational wave signals from binaries.

Our simulation setups lie near the transition between the wave-like and the particle-like regimes. Previous works on single non-spinning black hole simulations \cite{Traykova:2023letter} have shown a clear transition from Unruh accretion to Bondi accretion. The BBH system we studied exhibit a more gradual change, while the particle-like $\mu M=0.8$ case demonstrates a more distinct feature.

{Our study can easily be generalized in the following way}. We treated the scalar field as complex to simplify the force (avoiding rapid oscillations at the mass frequency $\omega = 2\mu$ as observed in Appendix C of Ref. \cite{Traykova:2021dua}]). A real scalar would induce additional high-frequency force oscillations at the Compton frequency, which could modulate the inspiral \cite{Traykova:2021dua}. We also neglected the backreaction of the scalar on the spacetime: in a fully coupled evolution the binary would inspiral self-consistently under both gravitational-wave emission and dynamical friction \cite{Bamber:2022pbs}. Including BH spin is another important generalization, since a spinning BH in a scalar medium experiences a transverse “spin-curvature” (Magnus) force \cite{Wang:2024cej,Dyson:2024qrq} (see also Ref. \cite{Kim:2024zjb} for magnetohydrodynamical Magnus effects). Finally, our simulations assumed an infinite homogeneous cloud. Realistic dark matter distributions (finite clouds or density gradients) and the binary’s history of passage through them would need to be accounted for in precise waveform modeling.
{Finally, our simulations have been carried out in a two-dimensional setting, akin to similar works in hydrodynamical accretion \cite{Lai:2022ylu}. It would be interested to systematically compare the importance of three-dimensional calculations in this simpler context. Also, we have primarily focused on the relativistic regime. It may be interesting to push this investigation to a sub-relativistic/large Bondi radius regime.}

In summary, this work provides a comprehensive relativistic analysis of dynamical friction and accretion drag on black hole binaries in a scalar field environment. We find that the presence of scalar dark matter consistently accelerates the merger (albeit the effect may be small under realistic astrophysical conditions), by extracting orbital energy. These results can be incorporated into gravitational waveform models as small environmental corrections. Our calculations thus serve as a stepping stone toward probing dark matter through gravitational waves \cite{chen2025muffled}.

\begin{acknowledgements}
    We are grateful to Lance Dixon for insightful discussions and thoughtful comments throughout the course of this work, and on the manuscript.
    Simulations were performed on DOE NERSC supercomputer Perlmutter under grants m4575 and m5801, which uses resources of the National Energy Research Scientific Computing Center, a DOE Office of Science User Facility supported by the Office of Science of the U.S. Department of Energy under Contract No. DE-AC02-05CH11231 using NERSC award NP-ERCAP0028480. 
    SX acknowledges support by US Department of Energy under contract DE-AC02-76SF00515. ERM acknowledges support by the National Science Foundation under grants No. PHY-2309210 and AST-2307394, as well as on the NSF Frontera supercomputer under grant AST21006, and on Delta at the National Center for
    Supercomputing Applications (NCSA) through allocation PHY210074 from the
    Advanced Cyberinfrastructure Coordination Ecosystem: Services \& Support
    (ACCESS) program, which is supported by National Science Foundation grants
    \#2138259, \#2138286, \#2138307, \#2137603, and \#2138296.
\end{acknowledgements}

\appendix

\begin{figure}
    \centering
    \includegraphics[width=0.9\linewidth]{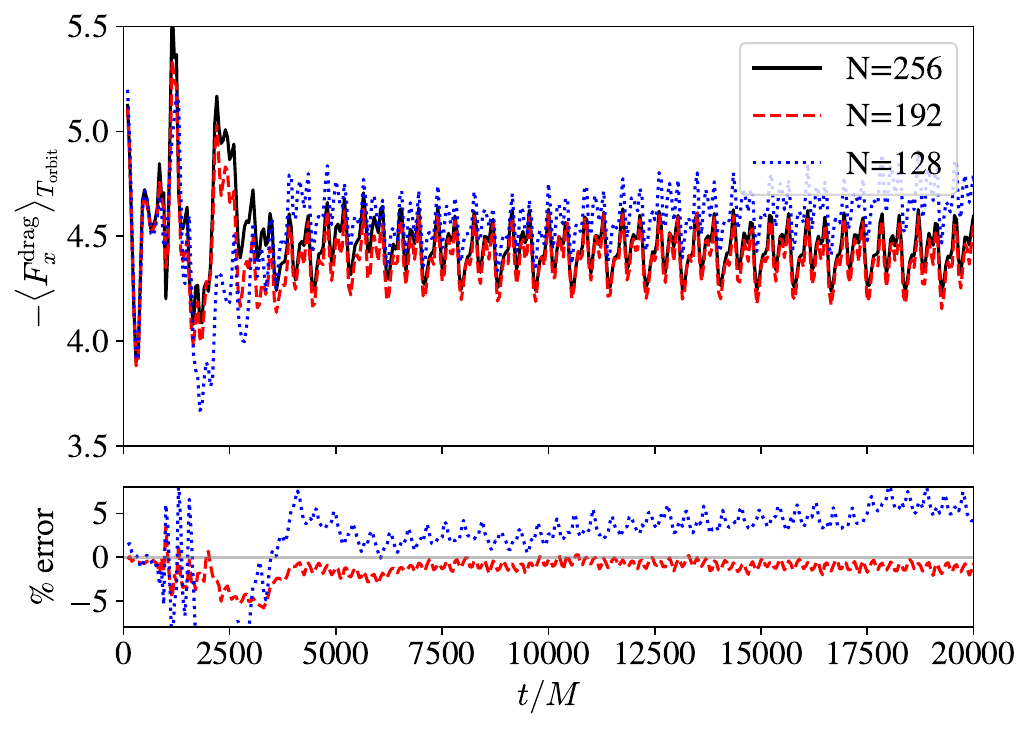}
    \caption{Convergence test: evolution of the drag force $F_x(t)$ for three grid resolutions ($N=128,192,256$).}
    \label{fig:n_cell_convergence}
\end{figure}

\begin{figure}
    \centering
    \includegraphics[width=0.9\linewidth]{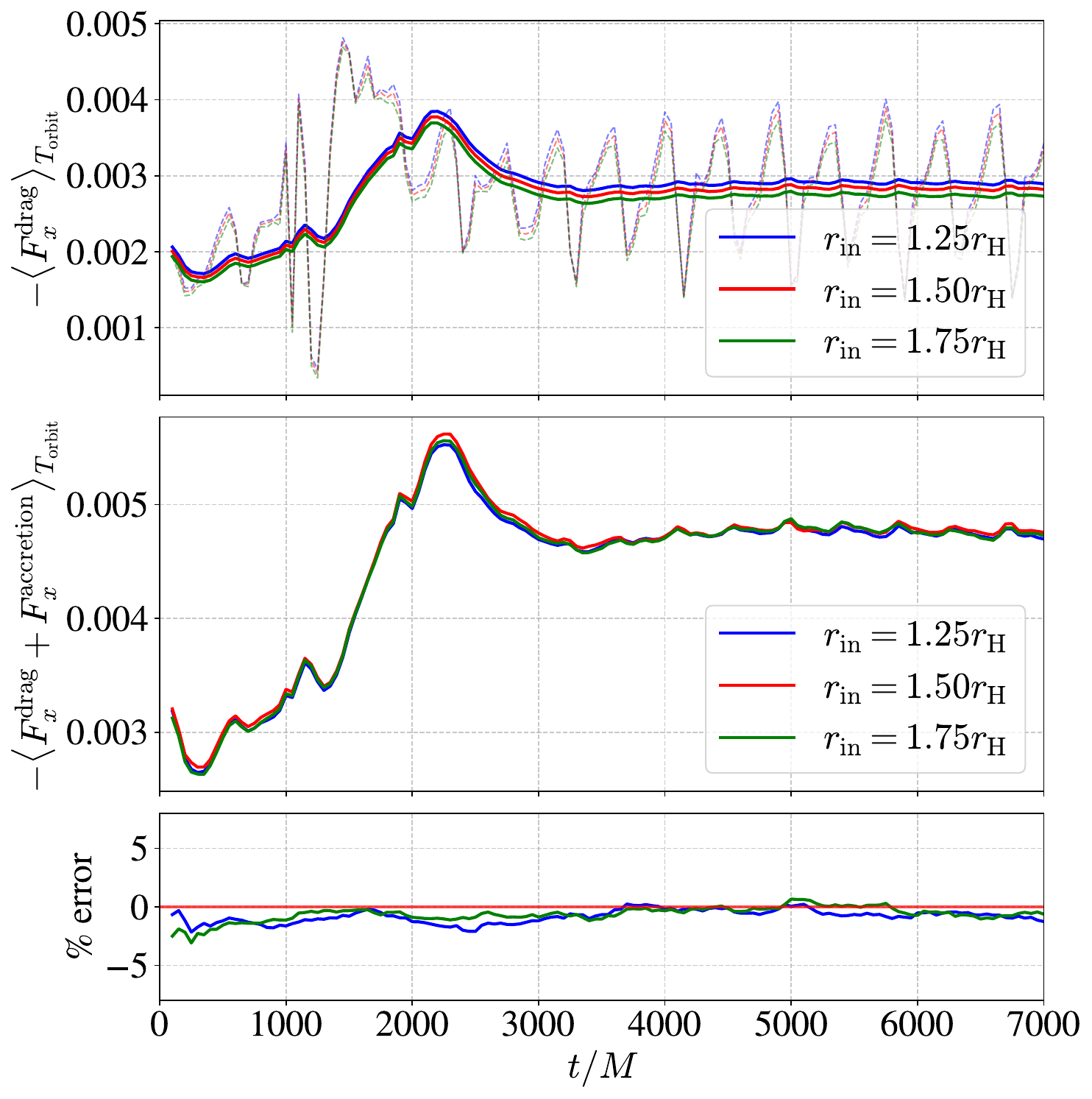}
    \caption{Upper panel: instantaneous and orbit-averaged drag force $F_x$ for the simulation with $v=0.5c,d_{\rm BBH}=26M,\mu M=0.2$ using three extraction radii $r_{\rm in}=1.25 r_{H},1.5 r_{H},1.75 r_{H}$. Dashed lines are the instantaneous volume drag $-F_x^{\rm drag}$ and solid lines are the orbit-averaged force. Middle panel: total force including both the volume drag and the horizon accretion flux. Bottom panel: fractional difference of each case relative to the baseline $r_{\rm in}=1.5 r_H$.}
    \label{fig:accretion_drag_rin_convergence}
\end{figure}

\begin{figure}[t]
    \centering
    \includegraphics[width=\linewidth]{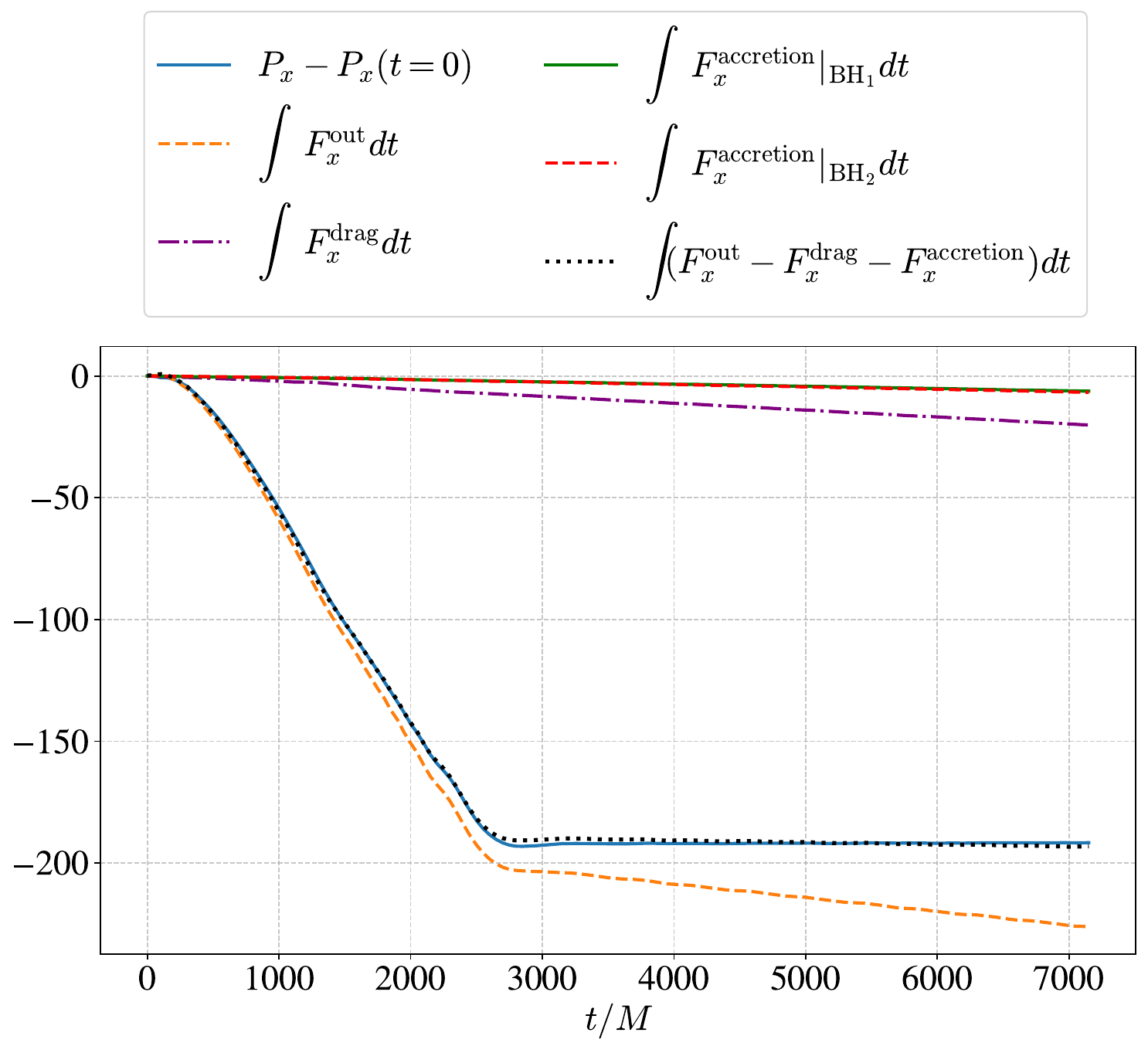} 
    \caption{Agreement between the time-integrated momentum balance terms for the simulation with $v=0.5c,d_{\rm BBH}=26M,\mu M=0.2$. The orange dashed line shows the outward flux $-\int F^{\rm out}_x\,dt'$, the purple dashed line shows the drag contribution $\int F^{\rm drag}_x\,dt'$, and the green/red lines correspond to the momentum accretion onto each BH. The blue solid curve is the change in total momentum change $\Delta P_x(t)$, and the black dotted curve is the sum of all flux and drag terms.}
    \label{fig:app_mom}
  \end{figure}
\begin{figure}
    \centering
    \includegraphics[width=\linewidth]{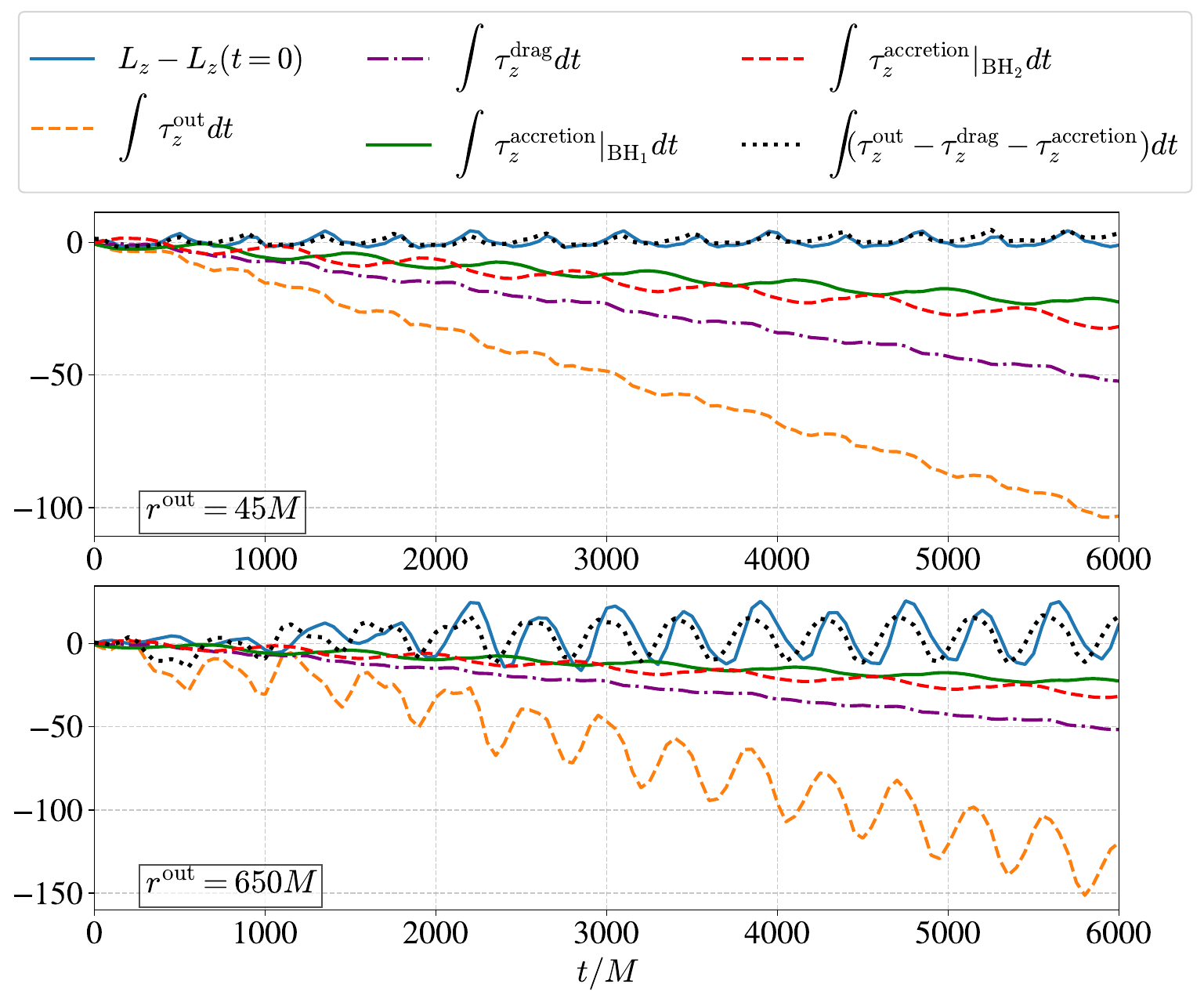} 
    \caption{Agreement between the time-integrated angular momentum fluxes. The orange dashed line shows the outward cumulative flux of $L_z$, the purple dashed  line is the cumulative drag torque, and the green/red lines are the $L_z$ fluxes accreted by each BH. The blue solid curve is the change in total angular momentum $\Delta L_z$, and the black dotted curve is the sum of all flux and friction contributions.}
    \label{fig:app_angmom}
\end{figure}

\section{Resolution and Convergence Tests}
\label{app:convergence}

{In order to provide an assessment of the numerical error of our results, we have performed a convergence study for a fiducial model in our study}. Figure~\ref{fig:n_cell_convergence} shows the results for an equal-mass, non-spinning binary in a light scalar field ($\mu M_{\rm tot}=0.05$) using three grid resolutions. The base grid uses cell spacing $\Delta x_{\rm base}$, with two finer grids at $\Delta x_{\rm base}/1.5$ and $\Delta x_{\rm base}/2$ (each with additional refinement). In Fig.~\ref{fig:n_cell_convergence} upper panel we plot the $x$-component of the force $F_x(t)$ on one BH over time for each resolution; the curves overlay closely, and differences shrink as resolution increases. The lower panel displays the difference between successive resolutions, scaled for second-order convergence. The nearly flat behavior of these scaled differences confirms that our code converges at roughly second order (small deviations at late times arise from interpolation error at refinement boundaries). We also monitor the total scalar-field energy (including outgoing radiation) and find it is conserved to better than 0.1\% in all runs. These tests validate the stability and accuracy of our implementation.

\section{Extraction Radius and Boundary Effects}
\label{app:boundary}

We define the force on each BH by integrating the scalar momentum flux through a spherical surface of radius $r_{\rm ex}$ around that BH. In principle, the true force is obtained as $r_{\rm ex}\to r_{\rm horizon}$, but in practice $r_{\rm ex}$ must be large enough to enclose the wave zone of the scalar field. In our production runs we used $r_{\rm ex}=5M$ (several times the horizon radius, $\approx0.5M$). To test sensitivity, we measured the force at $r_{\rm ex}=4M$, $5M$, and $6M$ for a representative BH (speed $v=0.5$, spin $a/M=0.7$). Figure~\ref{fig:accretion_drag_rin_convergence} shows the results. The upper panel plots the instantaneous and orbit-averaged drag force $F_x$ for the three choices of $r_{\rm ex}$. Dashed lines show the instantaneous volume-integral force $-F_x^{\rm drag}$, and solid lines show the orbit-averaged values. The middle panel plots the total force including both the volume drag and the horizon accretion flux. The bottom panel gives the fractional difference of each case relative to the baseline $r_{\rm ex}=5M$. We see that, beyond $r_{\rm ex}\approx4M$, all curves coincide within numerical uncertainty. This confirms that using $r_{\rm ex}=5M$ provides a robust measure of the force.

\section{Conservation}
\label{app:conservation}

\subsection{Momentum conservation and flux balance}
As a further validation, we monitor conservation of the total linear momentum in our simulation. We consider a control volume $\Sigma$ that excludes the two black hole horizons (bounded by inner excision surfaces and an outer spherical boundary at $r_{\rm out}$). The continuum balance law for the $x$-momentum in $\Sigma$ is
\begin{equation}
 -\frac{\partial}{\partial t}\int_{\Sigma} \sqrt{\gamma}\,S_x \,d^3x
 = -F^{\rm out}_{x} + F^{\rm drag}_{x} + F^{\rm acc}_{x,{\rm BH1}} + F^{\rm acc}_{x,{\rm BH2}},
\end{equation}
where $S_x$ is the $x$-momentum density, $F^{\rm out}_{x}$ is the flux of $x$-momentum through the outer boundary, $F^{\rm drag}_{x}$ is the dynamical friction contribution (integrated over the control volume sources), and $F^{\rm acc}_{x,{\rm BH}i}$ is the $x$-momentum accreted by $i$-th BH. Figure~\ref{fig:app_mom} plots each time-integrated term in this equation. The sum of all contributions (black dotted line) matches the actual change in total momentum $\Delta P_x$ inside $\Sigma$ (red solid line) to within numerical error, demonstrating that linear momentum is conserved.

\subsection{Angular momentum conservation}
We perform an analogous check for angular momentum around the $z$-axis. The total angular momentum $L_z$ inside $\Sigma$ satisfies a balance law, Eq.~\ref{eq_L_conserve}, with the drag torque and accretion torques replacing the linear momentum forces. In Fig.~\ref{fig:app_angmom} we plot the time-integrated outward flux of $L_z$, the drag torque, and the angular momentum accreted by each BH. The red curve shows the net change in total $L_z$ inside the volume. As in the linear case, the sum of the flux and torque terms (black dotted line) coincides with the actual change $\Delta L_z$ (red solid line), confirming that angular momentum is numerically conserved.

\bibliographystyle{unsrt}
\bibliography{refs}

\end{document}